\input{aipcheck}

\documentclass[
    ,final            
  ]
  {aipproc}

\def\simlt{\lower.5ex\hbox{$\; \buildrel < \over \sim \;$}}
\def\simgt{\lower.5ex\hbox{$\; \buildrel > \over \sim \;$}}
\layoutstyle{6x9}

\begin{document}

\title{Black hole masses and accretion states in ULXs}



\author{Roberto Soria}{
  address={MSSL, University College London, Holmbury St Mary, Dorking, Surrey, RH5 6NT, UK}
,altaddress={School of Physics, University of Sydney, NSW 2006, Australia}
}

\author{Zdenka Kuncic}{
  address={School of Physics, University of Sydney, NSW 2006, Australia}
}


\begin{abstract}
We summarize indirect empirical arguments
used for estimating black hole (BH) masses in ultraluminous
X-ray sources (ULXs). The interpretation of the X-ray data is still
too model-dependent to provide tight constraints,
but masses $\simlt 100 M_{\odot}$ seem the most likely.
It is getting clearer that ULXs do not show the same
evolutionary sequence between canonical spectral states
as stellar-mass BHs, nor the same timescale for state transitions.
Most ULX spectra are consistent either with a power-law-dominated
state (apparently identical to the canonical low/hard state),
or with a very high state (or slim-disk state).
Despite often showing luminosity variability,
there is little evidence of ULXs settling
into a canonical high/soft state, dominated
by a standard disk (disk-blackbody spectrum).
It is possible that the mass accretion rate
(but not necessarily the luminosity) is always larger than Eddington; 
but there may be additional physical differences between
stellar-mass BHs and ULXs, which disfavour transitions
to the standard-disk, radio-quiet state in the latter class.
We speculate that the hard state in ULXs is associated with jet 
or magnetic processes rather than an ADAF, can persist up 
to accretion rates $\approx$ Eddington, and can lead directly 
to the very high state.
\end{abstract}

\maketitle


\section{Indirect black hole mass estimates}

No kinematic (optical spectroscopic) mass measurements 
exist yet for black holes (BHs) in ultraluminous X-ray sources
(ULXs), although a program of phase-resolved VLT
observations for NGC\,1313 X-2 is currently under way
(Gris\'{e} et al., in preparation).
Indirect, model-dependent methods for estimating BH masses include:
\begin{enumerate}
\item assuming that the absorption-corrected
isotropic X-ray luminosity does not exceed
the isotropic Eddington luminosity of the BH:
$4\pi d^2 f_{\rm X} \simlt L_{\rm Edd} \approx 1.3 \times 10^{38}
\left(M/M_{\odot}\right)$ erg s$^{-1}$.
The estimated mass $M$ is only an upper limit
if we allow for moderate super-Eddington luminosity
or unisotropic emission (which is the case even
for a standard accretion disk without beaming).
Hence, a factor $\sim 1/2$--$1/4$ is sometimes implicitly 
allowed for, in the mass estimate;
\item using a standard disk model to relate the peak temperature
of a thermal spectral component to the innermost stable circular
orbit of the BH, and to the BH mass \cite{Maki:2000};
\item comparing caracteristic variability timescales in ULXs
with those in Galactic BHs and AGN, using the empirical
scaling $\tau \sim M$;
\item associating ULX spectral states with the "canonical"
BH state classification \cite{McR:2006};
the assumption here is that
transitions between states occur at fixed values or ranges of 
the self-similar accretion parameter $\dot{m} \equiv \dot{M}/\dot{M}_{\rm Edd}
\approx \left(0.1 c^2\dot{M}\right)/L_{\rm Edd}$.
Hence, $\dot{m} \sim (0.1/\eta) \left(L_{\rm X}/M\right)$, using $L_{\rm X}$ 
as a proxy for the accretion rate.
\end{enumerate}

Argument 1) suggests an upper limit of $\approx 100 M_{\odot}$ 
to the BH mass in ULXs, except perhaps for a handful
of sources \cite{Dehnen:2005} with $L_{\rm X} \approx 0.5$--$1.5 \times 10^{41}$, 
which may require intermediate-mass BHs or substantial beaming. 
A ULX in NGC\,1365 is a good example of a source 
that may reach its Eddington limit at $L_{\rm X} \approx 3 \times 10^{40}$ 
erg s$^{-1}$, triggering an outflow that blows away the accreting matter 
\cite{Soria:2007}. Such an upper mass limit is consistent 
with the maximum BH mass ($M \approx 70 M_{\odot}$) 
expected from individual stellar processes, i.e. direct collapse 
of a metal-poor, massive star \cite{Yungelson:2008}.

\section{Soft and hard states}
The other three arguments depend, more or less critically,
on whether ULX spectral states follow the canonical
classification for stellar-mass BHs. In this scenario, accreting BHs 
cycle between a low/hard state (power-law dominated, 
with photon index $\Gamma \approx 1.5$--$2$), a high/soft state 
(disk dominated, with an additional weak, steep power law 
with $\Gamma \approx 2.5$--$3.5$) and, occasionally, 
a very high state. The high/soft state is where Galactic BHs 
spend the majority of their time during their luminous 
phases ($0.1 L_{\rm Edd} \simlt L_{\rm X} \simlt L_{\rm Edd}$).  

Conversely, most ULXs are found in a state dominated by 
a broad, ``power-law-like'' component, 
with photon indices $1.5 \simlt \Gamma \simlt 3$; the index distribution 
peaks at $\Gamma  \approx 1.7$--$2$ \cite{Swartz:2004,Winter:2006}.
There is no gap or dichotomy between the flat and steep power-law 
sources. One or two additional features are sometimes present \cite{Stobbart:2006}:
i) a soft excess (probably direct thermal emission from a standard 
disk), usually at low temperatures ($0.1 \simlt kT_{\rm in} \simlt 0.2$ keV), 
which contributes only $\sim 10$--$20$ per cent of the X-ray luminosity; 
ii) a downward curvature or steepening of the spectrum at energies 
$\simgt 5$ keV. The presence or absence of those two components
in a given source often depends on the signal-to-noise
of the observation (hence, soft thermal components 
are more likely to be found in the brightest sources), 
and the amount of column density $N_{\rm H}$ 
which can mask the soft thermal component. No other properties
(such as spatial distribution or optical counterpart) have been 
found so far that can separate ULXs with a soft X-ray excess 
and/or spectral curvature from those in which those features 
are not detected.

Examples of ULXs in a classical high/soft state, dominated 
by a standard disk-blackbody spectrum, are very rare: 
only $\sim 10$ per cent of the sources with $L_{\rm X} > 10^{39}$ 
erg s$^{-1}$ in \cite{Swartz:2004}. The best-studied cases include 
M\,81 X-6 \cite{Swartz:2003}, and two sources in the colliding galaxies 
NGC\,4485/4490 \cite{Roberts:2002}. In all three cases, 
the fitted disk-blackbody temperatures are $kT_{\rm in} \approx 1$--$1.5$ 
keV, for an emitted luminosity $L_{\rm X} \simlt 4 \times 10^{39}$ erg s$^{-1}$.
Thus, those sources can easily be classified as the extreme end 
of the stellar-mass BH distribution. In other studies \cite{Winter:2006}, 
some ULXs have been assigned to the high/soft state because 
a disk-blackbody component was detected in their X-ray spectra: 
but crucially, in all those cases, the thermal 
component was relatively unimportant with respect 
to the power-law component, unlike the classical 
high/soft state of Galactic BHs.
Another group of ULXs, including many of those 
with $L_{\rm X} \approx 10^{40}$ erg s$^{-1}$ 
(e.g., IC\,342 X-1 \cite{Kubota:2002}, and a few more
discussed in \cite{Stobbart:2006}), can be fitted equally
well with a dominant disk-blackbody or similar thermal components, 
with $kT_{\rm in} \approx 1.5$--$2.5$ keV, or with 
a broken power-law or exponential cutoff above $\approx 5$--$7$ keV.
Again, this group of sources cannot be included 
in the classical high/soft state: their temperatures suggest 
that if their dominant X-ray emission component comes 
from the accretion disk, it must be heavily Comptonized, 
or in any case substantially different from the standard Shakura-Sunyaev spectrum 
\cite{SS:1973}. In the classical state classification, 
they belong more properly to the very high state 
(or perhaps the slim-disk state; see the next section) 
\cite{Maki:2007}. Finally, a handful of ``super-soft''
ULXs, (e.g., one in M\,101 \cite{Kong:2004} and another 
in NGC\,4631 \cite{Carpano:2007}), have a thermal spectrum 
with $kT \approx 70$ eV; there have been suggestions that 
they represent intermediate-mass BHs in the canonical high/soft state, 
but the most likely scenario is a transient, super-Eddington, 
nova-like source (a transient massive outflow from a white dwarf 
or from the accretion disk of a stellar-mass object).

Not only is it difficult to identify well-defined, canonical 
low/hard and high/soft states in the ULX population 
on a statistical basis, but it is 
also difficult to pinpoint transitions between  
canonical states in individual sources (unlike 
the case of stellar-mass BHs). For example, 
NGC\,1365 X-1 has been studied at various luminosity levels,  
varying between $L_{\rm X} \approx 3 \times 10^{39}$ erg s$^{-1}$ 
and $L_{\rm X} \approx 3 \times 10^{40}$ erg s$^{-1}$, 
but its spectrum is always dominated by a power-law with 
$\Gamma \approx 1.5$--$1.9$, with an additional 
(non-dominant) thermal component on some occasions \cite{Soria:2007}.
Clearly, this system is never in the high/soft state. 
But assigning its behaviour either to the very high state 
or the low/hard state is also problematic, given the large 
range of luminosities.

In summary, there is a clear lack of ULXs in a high/soft 
state (disk-blackbody component contributing $> 50$ per cent 
of the X-ray luminosity) with $3 \times 10^{39} \simlt L_{\rm X} 
\simlt 2 \times 10^{40}$ erg s$^{-1}$ {\emph {and}} 
$0.3 \simlt kT_{\rm in} \simlt 0.9$ keV (see also 
the luminosity-temperature plots in \cite{Miller:2004} 
and \cite{Feng:2005}).
That is the region of the parameter space where 
we would find accreting BHs with masses $\approx 30$--$100 M_{\odot}$ 
in their high/soft state, using the same 
standard-disk scaling and evolutionary track 
($L_{\rm disk} \sim T_{\rm in}^4$) 
that applies to Galactic BHs in that state.

Those findings suggest that {\emph {either}} BHs 
in that mass range do not exist (in contradiction 
with the Eddington-luminosity argument outlined earlier), 
{\emph {or}} they do not spend much time in the standard-disk 
dominated state. In the former scenario, the ULX population would consist 
either of beamed stellar-mass BHs, or of intermediate-mass BHs 
($M \sim 1000 M_{\odot}$). But both these possibilities are problematic.
Strong beaming seems inconsistent with the quasi-isotropic 
powering of ionized nebulae around some ULXs \cite{Kaaret:2004, 
Pakull:2008}; it is also inconsistent with quasi-periodic 
oscillations found in some ULXs, for example NGC\,5408 X-1 \cite{Stro:2007}. 
Intermediate-mass BHs require untested formation mechanisms, 
and are difficult to reconcile with the luminosity and spatial 
distribution of ULXs (more consistent with the upper end 
of high-mass X-ray binaries).

Therefore, we suggest that the lack of ULXs in that range of 
temperatures and luminosities is more likely due to the fact 
that they do not settle in a canonical high/soft state, and 
their X-ray spectral appearance is always heavily modified 
by some form of inverse Compton scattering.
In that case, the argument is then whether most ULXs belong 
to the low/hard state or the very high state 
(as we mentioned earlier, there is no evidence for separate 
softer and harder populations, for example a different 
spatial distribution or a gap in luminosities). 
The occasional presence of a faint, and sometimes low-temperature 
thermal component can be consistent with both states: 
in the low/hard state, the standard disk may be truncated 
at radii much larger than the innermost stable orbit; but 
a cooler, fainter disk component is also found 
in the very high state of Galactic BHs (the best example 
being XTE J1550$-$564, \cite{KD:2004, Soria2:2007}).
The Comptonizing corona is much hotter in the low/hard state, 
producing unbroken power-law spectra up to $\sim 100$ keV; 
therefore, ULXs with a spectral curvature or break at 
$\sim 5$--$10$ keV are more likely to be in the very high state, 
if we want to apply the canonical scheme.
The main reason why the low/hard and the very high state are easy 
to distinguish in Galactic sources is because the former occurs 
only at $L_{\rm X} \simlt$ a few per cent of $L_{\rm Edd}$ 
(when the accretion rate is thought to be too low  
to permit a disk-dominated state), 
and the latter at $L_{\rm X} \sim L_{\rm Edd}$ 
(when the accretion rate is too high for a standard disk).
If this were the case for ULXs, we would be forced to conclude 
that the sources in the very high state have masses 
$\simlt 100 M_{\odot}$ and those in the low/hard state 
have masses $> 1000 M_{\odot}$, as suggested in \cite{Winter:2006}. 
Instead, we speculate that in ULXs, in the absence of a canonical 
high/soft state, the low/hard state and the very high state 
can be contiguous, or perhaps even that there is 
no fundamental qualitative difference betwen the two, so that 
sources can appear to have a ``low/hard'' spectrum even 
for luminosities approaching the Eddington limit.

\section{Where is the accretion disk?}
There is no lack of physical and phenomenological models 
to explain why the X-ray spectrum cannot be dominated 
by a standard disk at accretion rates $\dot{m} \simgt 1$. 
The presence of non-dominant, relatively cool disk-blackbody 
emission in addition to a stronger, broader power-law-like 
component can be interpreted qualitatively as a two-phase
structure: a standard disk directly visible outside a transition radius $R_{\rm c}$,
and a "modified" (e.g., heavily Comptonized) disk or flow at $R < R_{\rm c}$.
Typically, the broader, dominant spectral component accounts for $\sim 90$ per cent 
of the radiated power, hence we expect $R_{\rm c} \sim 50$--$100 R_{\rm g}$ 
$\sim 10 R_{\rm ISCO}$. The fitted temperature of the soft component 
(outer disk) is then expected to be $kT_{\rm in} \approx kT(R_{\rm c}) 
\approx  kT(R_{\rm ISCO})(R/R_{\rm ISCO})^{-3/4} \sim 0.1$--$0.3$ keV 
for $M \sim 10$--$100 M_{\odot}$.  (If $L_{\rm X} \approx 10^{40}$ erg s$^{-1}$, 
masses nearer the upper limit of this range are also more consistent 
with the Eddington-luminosity argument).

This simple argument can explain the characteristic temperature and luminosity 
of the soft excess found in many ULXs without invoking intermediate-mass BHs.  
More importantly, there is evidence that this is what happened in the Galactic BH 
XTE J1550$-$564 ($M \approx 10 M_{\odot}$) when it reached the very high state 
in the 1998 outburst \cite{KD:2004, KM:2004, DK:2006, Soria2:2007}.
In that phase of the outburst, the spectrum became dominated by 
a power-law-like component; the fitted peak temperature 
of the disk-blackbody component decreased from $\approx 1$ keV 
(as expected for a BH of this mass) to $\approx 0.4$ keV, but without 
a decrease in the fitted disk-blackbody luminosity\footnote{The 
reliability of the fitted disk-blackbody temperatures in 
the RXTE/PCA data from the 1998 outburst, originally presented 
in \cite{Sobczak_sp:2000}, has sometimes been questioned or dismissed: 
e.g., J. McClintock, priv.~comm.; T.~Belloni, priv.~comm. 
On the other hand, independent re-analysis of the data 
by C.~Done and collaborators has confirmed the low temperature values. 
In our opinion, the simultaneous correlated increase in the observed 
QPO frequency over the same days provides convincing evidence 
that both sets of measurements are reliable and are tracing 
a real physical effect.}. This finding can be explained if the 
disk-blackbody component has an increasing (receding) 
inner radius $R_{\rm c}$ at increasing accretion rate $\dot{m}$. 
Crucially, during that same days, a low-frequency QPO 
was detected \cite{Sobczak_tm:2000}, 
with a varying frequency, inversely correlated 
with the fitted inner-disk radius \cite{Soria2:2007}. 
This is consistent with the QPO frequency being related 
to the transition radius $R_{\rm c}$ between the outer standard disk 
and the inner Comptonizing region, and provides further evidence 
in support of a receding $R_{\rm c}$ at high accretion rates.
There are various scenarios or physical models 
that can explain a transition between standard disk and non-standard 
inflow, with a transition radus moving outwards at increasing $\dot{m}$.
Some of them are:
\begin{enumerate}
\item the inner disk is covered or partly replaced 
by a warm corona with $kT \sim$ a few keV 
(much cooler than in the low/hard state, 
leading to a noticeable break at energies $\simgt 5$ keV),
and scattering optical depth $\tau^{\rm es} \sim$ a few \cite{DK:2006, Goad:2006, 
Stobbart:2006, Roberts:2007}. $R_{\rm c}$ is interpreted as the outer boundary 
of the corona, beyond which the geometrically-thin standard disk is visible. 
This scenario requires that most accretion power be dissipated 
in the corona rather than the inner disk, or be transferred efficiently 
to the corona rather than directly radiated. A related version of 
this scenario was proposed by \cite{Socrates:2006}, with the inner disk 
covered by a hotter ($kT \sim 100$ keV), moderately 
optically-thin ($\tau^{\rm es} \simlt 1$) corona, 
and high albedo at the photoionized inner-disk surface;
this, and the dissipation of most accretion power inside the corona, 
ensures a non-thermal outgoing X-ray spectrum from the inner region.
This model is more suitable for ULX spectra with a soft excess 
but no high-energy breaks; for example those that may have been 
classified in the low/hard state, if we used canonical states.

\item a physical model related to the previous coronal scenarios is based 
on the standard disk itself becoming hotter, effectively thin
($\tau^{\rm eff}_{\nu} \approx \sqrt{\tau^{\rm ff}_{\nu}\tau^{\rm es}} \simlt 1$)
but still thick to scattering ($\tau^{\rm es} \sim 10$) 
in the inner region, when $\dot{m} \simgt 1$ 
\cite{SS:1973, Callahan:1977, Elvis:1987, Belo:1998, Kawaguchi:2003, Artemova:2006}.
It was already known since \cite{SS:1973} that in the radiation-pressure-dominated 
zone of the disk, the electron density $n_e \sim \dot{m}^{-2}$; this is why 
the inner part of the disk eventually becomes effectively thin at high  
accretion rates. When that happens, the inner disk radiates less efficiently 
than a blackbody at a given temperature, so this is compensated 
by an increase of the temperature, up to $\sim$ a few keV. 
The outgoing spectrum becomes heavily Comptonized, with 
a power-law-like appearance in the $2$--$10$ keV band. 
Unlike the corona models, here there is no need 
to invoke a separate physical object covering or replacing the disk: 
it is the inner disk itself that morphs into a geometrically-thick, 
scattering-dominated 
region, with physical parameters consistent with those required 
for ULX spectral fits. In our phenomenological application to ULXs, 
$R_{\rm c}$ can be identified with the thick/thin transition radius. 
A self-similar analytic approximation from \cite{SS:1973} suggests
$R_{\rm c}/R_{\rm ISCO} \approx 12 \alpha^{34/93} m^{2/93} \dot{m}^{32/39} 
\left[1-\left(R_{\rm c}/R_{\rm ISCO}\right)^{-1/2}\right]^{64/93} \sim 
{\rm a\ few\ \ } \dot{m}^{0.69}$, where $m$ is the BH mass in solar 
units\footnote{The numerical coefficient in front of this expression 
differs from the one in \cite{SS:1973} because we are using a higher 
free-free absorption coefficient, suitable to cosmic abundances; 
see also \cite{Frank:2002}.}. Hence, the transition between standard 
disk and Comptonizing region appears when the accretion rate approaches 
Eddington; moves outwards for increasing $\dot{m}$ (in agreement with 
the very-high-state interpretation of XTE J1550$-$564 and some ULXs); 
and the effect is stronger for a high viscosity parameter $\alpha$;

\item at super-Eddington accretion rates ($\dot{m} \simgt 2$), 
the inner region of the accretion flow becomes a ``slim disk''\cite{Abra:1988}: 
an optically-thick solution with energy advection and $H/R \sim 1$.
Characteristic slim-disk temperatures are $kT_{\rm in} \approx 1.5$--$2.5$ keV, 
with $L_{\rm disk} \sim T_{\rm in}^2$; the radial 
temperature profile in the disk is flatter than $R^{-3/4}$. 
The emitted luminosity saturates at $\sim$ a few $L_{\rm Edd}$ 
for $\dot{m} \sim 10$, partly because of photon trapping;
the transition between outer standard disk and photon-trapping region 
occurs at $R_{\rm c} \sim (\dot{m})^2 R_{\rm g}$ \cite{Ohsuga:2003, Ohsuga:2005}.
In this scenario, the slim disk provides the dominant broad
component with an exponential cutoff above $\sim 5$ keV.
It predicts similar spectra to those expected 
from Comptonization in a warm, thick corona \cite{Watarai:2001, Maki:2007}.
Hence, it is more suitable for ULXs that show strong 
spectral curvature (which can also be modelled with 
a dominant disk-blackbody component 
at $kT_{\rm in} \approx 1.5$--$2.5$ keV, e.g., \cite{Stobbart:2006}); 
for example, IC 342 X-1 \cite{Kubota:2002, Ebisawa:2003}.
In the slim-disk scenarios, BH masses in ULXs are 
$\sim 30$--$100 M_{\odot}$. When the spectral curvature 
is less prominent, or occurs $> 5$ keV, Comptonization models 
may be more suitable \cite{Maki:2007};

\item in addition to the thick/thin transition discussed earlier, 
standard disks have another characteristic radius that becomes 
relevant for $\dot{m} \simgt 1$: the spherization radius 
$R_{\rm sp} \sim  (9/4) \dot{m} \, R_{\rm ISCO}$. 
At $R \approx R_{\rm sp}$,  
the thin-disk approximation breaks down ($H/R \simgt 0.5$) 
because radiation pressure dominates over gravity; strong outflows 
are launched at $R \simlt R_{\rm sp}$. The total disk luminosity 
is $\approx L_{\rm Edd} \times (1 + \ln \dot{m})$. Of this, 
$L \approx L_{\rm Edd}$ is released at $R>R_{\rm sp}$ 
and $L \approx (\ln \dot{m}) \, L_{\rm Edd}$ in the outflow 
region; the latter component can also be mildly beamed 
by the outflow itself, if seen face-on.
It was suggested \cite{Poutanen:2006, King:2008} that 
the spherization radius corresponds to the characteristic 
transition radius $R_{\rm c}$ between outer (standard) disk 
and inner inflow, required for most ULX models. 
For accretion rates $\dot{m} \sim 10$--$100$ and 
stellar-mass BHs, the model reproduces the characteristic temperature 
of the soft excess ($kT(R_{\rm sp})$), the total luminosity and the relative 
contribution of the inner and outer regions. However, it may have more 
difficulties in explaining the power-law-like (or cutoff power-law) spectrum 
of the emission from the inner region. Also, the low intrinsic $N_{\rm H}$ fitted 
to most ULX spectra suggests that they are not seen through strong outflows;

\item finally, the centrifugal boundary layer model \cite{CT:1995}
may be a promising (and so far, not fully exploited) 
tool to understand some ULX spectra. The boundary layer 
is caused by an adjustment of the Keplerian disk to 
the sub-Keplerian boundary conditions near the BH; 
its position depends on $\dot{m}$ and viscosity, 
through the Reynolds number \cite{Titarchuk:1998}.
The boundary layer consists of standing or oscillating 
shock waves that accelerate electrons very efficiently, producing a non-thermal 
(power-law) spectrum \cite{Mandal:2005}, in addition to the seed  
disk-blackbody component from the standard outer disk, before 
the shock. This model predicts testable correlations between QPO frequencies 
(oscillations of the bounday layer) and photon index of the 
power-law component. 

\end{enumerate}

\section{Conclusions}

X-ray spectral and timing studies have not yet provided
a tight constraint on ULX BH masses, because they are still 
too model-dependent.
The simplest Eddington-limit argument suggests $M \simlt 100 M_{\odot}$ 
for all but a handful of ULXs; it remains the least controversial, 
and is also in agreement with most spectral models.

The most significant finding from X-ray spectroscopy 
is that ULXs are much less likely than Galactic BHs 
to settle in the canonical high/soft state, dominated 
by a standard disk. Some ULX spectra are dominated
by a hard power-law (apparently identical to canonical 
low/hard state spectra); others have a softer (steeper) 
spectrum, sometimes with a soft thermal excess and 
a high-energy downturn, consistent with heavily Comptonized 
emission from a standard disk, or a slim disk.

There are various plausible models that may explain 
the luminosity and spectral shape for accretion rates $\dot{m} \simgt 1$.
However, many ULXs show flux variability by at least a factor of a few; 
some are transients, so we know that the accretion rate is sometimes 
strongly reduced or switched off. We would expect to find some of them, 
at some epochs, in a canonical high/soft state, when $\dot{m} \sim 0.1$--$1$.
This would also give us a chance to identify their 
standard $L_{\rm disk} \sim T_{\rm in}^4$ track and better constrain their mass.
But this is not the case, although some ULXs have been seen to switch 
between a hard power-law state and a steeper spectrum 
with high-energy curvature. 
Furthermore, ULXs with a pure hard power-law spectrum are not 
significantly less luminous than those with a softer spectrum 
or a soft excess; some of them also exceed $10^{40}$ erg s$^{-1}$. 
This is unlike the canonical low/hard state in Galactic BHs.

We speculate that there must be a fundamental
physical difference between Galactic BHs and ULXs, which 
prevents the latter from settling into 
a long-duration disk-dominated state. For example,
ULXs may switch between harder and softer states 
depending on the optical depth and temperature of 
the Comptonizing region but would always be dominated 
by Comptonized emission. If ULX BHs have masses $\sim 30$--$100 M_{\odot}$, 
it may seem unlikely that such a small mass difference 
may suppress the disk-dominated state. After all, 
standard disks are seen in AGN with masses $\sim 10^6$--$10^9 M_{\odot}$.
The type of donor star (probably OB stars in ULXs) may have an effect 
on the duty cycle, keeping them in a bright state for longer periods 
of time than soft X-ray transients (powered by low-mass giants).
But it is not clear why it would have an effect on the disk stability.

Perhaps the key is in the nature of the low/hard state.
If/when it consists of a truncated disk replaced by a radiatively-inefficient, 
advection-dominated flow, it has to be limited to accretion rates 
$\dot{m} \simlt 0.01$. But it is likely that in some Galactic BHs, 
the low/hard state has a fully-formed disk with a jet \cite{Miller:2006}. 
If so, the key element that defines the low/hard state is that most 
of the accretion power is carried out non-radiatively, in a jet, wind 
or Poynting flux \cite{Kuncic:2004}.
The transition from the low/hard to the high/soft state would correspond 
to the suppression of the jet or Poynting flux, with ejections or flaring 
resuming as the source enters the very high state. (We do not know 
whether ULXs also possess a radio-quiet state, or it is suppressed 
together with the high/soft state).
Therefore, we speculate that ULXs can remain in a power-law dominated state 
(similar to the low/hard state) up to $\dot{m} \sim 1$ and
then switch directly to a very high state or outflow-dominated 
or slim-disk state, as the accretion rate (but not necessarily 
the luminosity) increases above Eddington.



\end{document}